\documentclass[12pt,fleqn]{article}

\usepackage{amssymb}
\usepackage{epsf}

\begin{document}
\title{A New Atom Trap: The Annular Shell Atom Trap (ASAT)} 

\author{Herschel S. Pilloff $^{1}$ and Marko Horbatsch $^{2}$}
\date{February 12, 2002}
\maketitle
\noindent$^{1} $Physics Department, University of Maryland Baltimore County, Baltimore, MD 21250 
\\ \hspace*{.25in} email: hersch\_pilloff@umbc.edu
\\$^{2} $Department of Physics and Astronomy, York University, Toronto, Ontario M3J 1P3, Canada
\\ \hspace*{.25in} email: marko@yorku.ca    

\begin{center}
{\bf Abstract}
\end{center}

In the course of exploring some aspects of atom guiding in a hollow, optical fiber, a small negative potential energy well was found just in front of the repulsive or guiding barrier.  This results from the optical dipole and the van der Waals potentials.  The ground state for atoms bound in this negative potential well was determined by numerically solving the Schr\"odinger eq. and it was found that this negative well could serve as an atom trap.  This trap is referred to as the Annular Shell Atom Trap or ASAT because of the geometry of the trapped atoms which are located in the locus of points defining a very thin annular shell just in front of the guiding barrier.  A unique feature of the ASAT is the compression of the atoms from the entire volume to the volume of the annular shell resulting in a very high density of atoms in this trap.  This trap may have applications to very low temperatures using evaporative cooling and possibly the formation of BEC.  Finally, a scheme is discussed for taking advantage of the de Broglie wavelength to store atoms in a bottle trap based on the inability of long de Broglie wavelengths to escape through a selective de Broglie wavelength filter in the atom bottle trap.

\vspace{0.1in}

\vspace{0.1in}

\newpage
\small\normalsize

The polarizability of an atom is almost always positive but, it can be negative and some unusual effects can then be observed.  This can occur when a laser or other monochromatic source is tuned slightly above or to the ``blue'' of an atomic resonance. The interaction of the external field on the atom through its negative polarizability produces a gradient dipole force which tends to drive the atom to regions of minimum intensity. Cook and Hill\cite{cook} suggested using an evanescent wave to produce an atom mirror outside of a dielectric.  Zoller, et.al.\cite{marksteiner, savage} analyzed the case for a clad, hollow fiber in which the external field was confined to the annular region and used the resulting evanescent field in the hollow region to guide atoms. In what has become known as ``blue-guiding", Renn\cite{renn} and Ito\cite{ito} have experimentally demonstrated evanescent wave guiding of rubidium atoms in hollow optical fibers.  A comprehensive review of evanescent light-wave atom optics has been published by Dowling and Gea-Banacloche\cite{dowling}.

Atom guiding in a hollow, metal-coated optical fiber has previously been analyzed\cite{pilloff},\cite{pilloff1}.  The metal coating is assumed to be a perfect conductor and this provides both simpler boundary conditions for determining exact solutions of the $TM_{0n}$ modes and maximizes the evanescent guiding field for a given laser electric field injected in the annular region of the fiber.  
A typical potential for this interaction is shown in Fig.1 and includes both the quantum optical dipole (odp) and the van der Waal's (vdW) potentials.  Here 

\begin{equation}
U_{odp}=\frac{1}{2}{\it \hbar}\,\Delta\,\ln (1+2\,{\frac {{d^{2}E{^2}}}{{{\it \gamma}
}^{2}+4\,{\Delta}^{2}}})
\end{equation}
where $\Delta=\omega - \omega_{o} >0$ is the detuning,  $\omega$ the laser frequency, $\omega_{o}$ the resonance frequency of the two level atom, $\gamma$ the decay rate of the  upper level, $d$ the transition dipole moment between levels 1 and 2, and $E$ the evanescent electric field amplitude in the hollow region of the fiber as calculated from the Helmholtz equation.  The vdW potential is given by

\begin{equation}\label{U_vdw}
U_{vdW}=\frac{-1}{32\,\pi \epsilon_{0}} \left(\frac{n_2^{2} -1}{n_2^{2} +1}\right)\left(\frac{1}{(a-r)^{3}}-\frac{1}{a^{3}}\right) d^{\,2}
\end{equation}

\noindent where $\epsilon_{0}$ is the vacuum permittivity, $n_{2}=\sqrt5$ is the dielectric constant of the annular region of the fiber, $a$ the inner radius of the fiber, and  $r$ the distance from the center to the atom.  The total potential is given by $U_{tot} = U_{odp} + U_{vdW}$.  Note that a synthetic atom has been used here where the resonance wavelength corresponds to 5000 A${\!\!\,^{^{o}}}$ and the atom mass, $m_0$, is $4.00 \times 10^{-26}$ kg.
Further details can be found in\cite{pilloff}, \cite{pilloff1} and references therein.  

While Fig.1  is familiar because it shows the repulsive or guiding barrier in front of the inner wall, a new result is obtained when the energy resolution or magnification is increased by $\simeq \times 10^4$.  In this case a small, negative potential energy well is observed in Fig.2.  This is a result of the vdW potential dominating both 
very far from and
very close to the inner wall whereas the odp potential dominates in the region where the rising leading-edge of the pulse in Fig.1 has a strongly positive slope.  
The reason for the strongly positive slope is the very fast "turn-on" of the odp potential here.
At the point where the potential has a maximum, the slope quickly becomes strongly negative and the vdW potential becomes dominant again.   Any atoms on this side of the barrier will quickly be drawn into the wall.
These figures display 1-dimensional potentials which are radially symmetric but the energy axis in each figure has a different scale.  A cross section through the repulsive barrier is shown  in Fig.3, whereas the negative potential energy well is too shallow to be seen on this energy scale.

When used with a single wavelength odp potential as in eq.$\!$ 1, the depth of the negative potential well is comparatively shallow.  However, using a two-color odp potential can significantly increase the well 
depth\cite{ovchinnikov},\cite{mabuchi}.
The two-color total potential is:

\begin{equation}
U_{2\_color}=U_{odp}(\Delta_{1},E_{1}) + U_{vdW} + U_{odp}(\Delta_{2},E_{2})
\end{equation}
where the detunings are $\Delta_1 >0$ and $\Delta_2 <0$, and $E_1$ and $E_2$ are the electric field amplitudes in the hollow region for the corresponding laser wavelengths $\lambda_1$ and $\lambda_2$, and both wavelengths are injected into the annular region of the hollow fiber. 
Calculations have shown that the negative well depth can be increased by factors up to $10^4$.
 Note that $\Delta_{2}$ is negative and tuned to the red of the transition frequency.

The 2-color odp potential can be explained using the Red-Blue Pushme-Pullyou resonator or 
trap$\cite{dowling}$.  Blue detuned evanescent radiation pushes atoms away from the wall and toward regions of lower intensity while red detuned evanescent radiation pulls atoms toward the wall and higher intensity regions.   By careful external focusing, it is possible to achieve a spatial separation between attractive and repulsive regions  in the case of the dielectric slab$\cite{ovchinnikov}$ or by using a 3-level atom in the quartz microsphere$\cite{mabuchi}$.  It is also possible to obtain a useful spatial separation between the blue and red-detuned light for the two-level atom used here.  

While both wave lengths are injected into the annular region of the fiber, the two parameters, $E_i$ and $\Delta_i$, in eq. 3  determine the characteristics of each evanescent wave.  For example, for the red detuned evanescent wave, the radial dependence of the amplitude can be adjusted to drop (a negative potential here) before the blue-detuned wave eventually rises for larger values of r.  In moving toward the inner wall, the atom  encounters a region containing a red detuned wave and then, for increasing values of r, enters a different region where the blue detuned wave is also present.  As the red odp together with the vdW potential attracts the atom toward the repulsive barrier, the blue radiation repels the atom as it approaches the large energy barrier.  At some point there is a superposition of forces such that the repulsive force due to the blue light competes against the attractive forces of both the red odp plus the vdW potential.  If the combined attraction of the latter is greater than the repulsive blue force, the atom will continue to move closer toward the inner wall and the well will deepen - vice versa and the well becomes shallower.  While the blue odp potential is usually of larger magnitude, it is important that the combined red forces are first encountered at smaller r values in moving toward the wall so that the negative well depth can be enhanced.

Consider an ensemble of atoms inside the hollow fiber.  These atoms randomly interact with the repulsive barrier just in front of the inner wall.  Since these interactions are assumed to be elastic, the atoms must lose energy by collisions with other atoms through the process of evaporative cooling\cite{masuhara} and  many of the remaining atoms will eventually be trapped in the negative potential well.  At high laser powers, the trapped atoms are expected to have a broad distribution of energies.  Because both the negative well and the repulsive barrier are extremely sensitive to reducing or varying the laser power and/or the laser detuning as shown in
 Fig.4.  In this case, using eq. 3, $\Delta_{1}$ is adiabatically varied with time, and a series of time-frozen snapshots are shown for $\Delta_{1i}$ where $i=1...5$.  All other parameters $(E_{1}$, $E_{2}$, and $\Delta_{2})$ are held constant.
This can provide an additional level of control in that the well can become deeper, and move toward the inner wall, while the barrier eventually becomes negative and may be very thin where it actually exists.  The hotter atoms escape from the well preferentially to the inner wall because the barrier has been eliminated or because atom tunneling through what remains of the barrier provides a significant loss mechanism.  By reducing the barrier adiabatically, the evaporative cooling can be made very efficient and the loss of atoms from the trap can be minimized.

Before calculating the ground-state energy for atoms in this trap, it should be noted that  
the ground-state eigenvalue should be as deep as possible.  The selectivity of this system is complex.  Following de Broglie's approach as described by Bohm\cite{Bohm} and modifying it to use a stationary wave to represent an atom in a stationary state requires that the stationary wave must fit continuously on itself after going around the radial potential well inside the hollow fiber.  This requires an integral number of waves fitting in the above described circumference.  The atoms must be very cold and since their de Broglie wavelength, $\lambda_{dB}$, must fit within the circumference of the hollow fiber, it may be quite long.  For example, for 
$a = 10 \mu$m, then as determined by de Broglie's constraint,  $\lambda_{dB} \le 20 \pi $ $ \mu$m.  
But the low temperatures desired may give $\lambda_{dB}$ which would be even longer and therefore can't fit within the circumference of the hollow fiber and is not allowed. The conclusion is that the minimum eigenvalue depends on the inner radius of the hollow fiber.  Larger inner radius fibers sustain lower eigenvalues (as well as a higher number of states) in the negative energy well and this effect has been observed in the numerical calculations.

In order to obtain a numerical solution of Schr\"odinger's eq., it is helpful to write it in dimensionless form by letting $\rho=r/a$.  The result is the scale factor, SF, given by 
\begin{equation}
SF=2 m_o \displaystyle\frac{a^2}{\hbar^2}
\end{equation}

For the case where there is no azimuthal dependence, then the radial Schr\"odinger eq. becomes:

\begin{equation}
{\frac {d^{2}}{d{\rho}^{2}}}R(\rho)+\frac{1}{\rho}\frac{d}{d\rho} R(\rho) +
\displaystyle\frac{2m_o a^2}{\hbar^2}(E-U(r))R(\rho)=0.
\end{equation}
where $R(\rho)$ is the radial wave function.

The dimensionless radial Schr\"odinger equation was solved using Maple with
a shooting algorithm based on a stiff ordinary differential equation
solver (Gear's method). The potential $\left(-\displaystyle\frac{2 m_o a^2}{\hbar^2}\right)U(r)$ has a large positive barrier 
at $\rho < 1$, and, thus, the problem can be solved by approximating the
long-lived resonance problem as a bound state. The bound-state problem
is defined by setting a boundary condition such that the wavefunction vanishes
inside this potential barrier.

The results for the eigenenergy were confirmed using three approaches
that determine the eigenvalue by a trial method using a numerical
initial-value problem solver:
{\it (i)} a shooting algorithm that starts the solution at $\rho=0$ and ensures
that the solution vanishes at $\rho=\rho_1$ without crossing the $\rho$-axis;
{\it (ii)} the same algorithm which starts the solution at $\rho=\rho_1$, and
which propagates the solution inwards;
{\it (iii)} a matching algorithm which combines information from both
inside-out and outside-in integrations, and which is deemed to be numerically
more stable. 
Condition {\it (iii)} is shown below as:
 
\begin{equation} 
\frac
{\psi^ \prime(\rho)}{ \psi(\rho)\:}{\displaystyle|_{\rho \rightarrow \rho_{\rm match}^-}}
-\frac{ \psi^ \prime(\rho)}{ \psi(\rho)\:}{\displaystyle|_{\rho \rightarrow \rho_{\rm match}^+}}=0
\end{equation}
\noindent
where ${\displaystyle|_{\rho \rightarrow \rho_{\rm match}^-}}$ and 
${\displaystyle|_{\rho \rightarrow \rho_{\rm match}^+}}$ represent $\rho$ propagating from left to right, and right to left, respectively, to the matching point.

The sensitivity of the eigenvalue to the choice of the right-hand
boundary $\rho_1$ where the boundary condition was set to $\psi(\rho_1)=0$ was explored.  A range was found
for $\rho_1$ values such that the eigenvalue remained constant.

The determined ground-state energy eigenvalue was verified by increasing the
numerical precision. It can be observed from Fig.5 that the potential well
is quite narrow, and that the Heisenberg uncertainty principle prevents the
atoms from accumulating near the classical potential energy minimum. For this
potential, the atoms are bound with an energy that corresponds to
about one quarter of the value of the potential at the minimum, and the atoms
spend a considerable amount of time inside the potential barrier.
The results are shown in Fig.5 where the ground state energy is -25,668 and corresponds to $3.54\times 10^{-31}$ J.  The counter propagating solutions are matched at $\rho = 0.97$.

An interesting feature of this trap is the increased atom density that results from the compression of 
the atoms in the original volume to the much smaller volume of the annular shell.  Consider the following simple model for atom compression.  A hollow fiber of unit length is initially filled at some pressure which is proportional to 
$\displaystyle\frac{1}{\pi a^2}$.  These atoms are compressed and the new pressure is inversely proportional to the area of the annulus.   This can be written as
$\displaystyle\frac{1}{\pi (r_2^2 -r_1^2)}$ where $r_2$ and $r_1$ are the outer and inner radius of the annulus, respectively.  Writing this as $\displaystyle\frac{1}{\pi (r_2+r_1)(r_2-r_1)}$, this can be approximated as 
$\displaystyle\frac{1}{\pi (2 r_{ave})(\delta_{r})}$ where $r_{ave}$ can be approximated by the location of the minimum in the potential and $\delta_r$ is the width of the annulus.  Since the minimum in the potential occurs about $0.15 \mu$m from the inner wall, $r_{ave} \approx a$.
The distance between the classical turning points is a good estimate for $\delta_r$ but this would be a low 
estimate whereas $\sim1.5$ times this value would account for tunneling which occurs mostly on the inside. The compression factor, $C$, becomes 
\begin{equation}
C \approx \displaystyle\frac{a}{3\, \delta_{r}}
\end{equation}
where in this simple model, the compression is assumed to be lossless.
As an example, consider a hollow fiber where 
$a=10^{-2}$ m and $\delta_{r} = 10^{-7}$ m.  This gives $C= 3.33 \times 10^{4}$ which considering likely losses still seems quite significant.

The characterization of this trap particularly as it involves the details of evaporative cooling is important for several applications such as the limiting temperature that can be achieved and whether BEC might be easily achieved in this trap.  An introduction to these issues is discussed by Metcalf \cite{metcalf}, but these are beyond the scope of the present paper\cite{pilloff2}.  Also, it is expected that the negative potential energy well will also exist for a slab geometry and for the ``atom trampoline" in which an atom in a gravitational field will bounce up and down on top of the dielectric slab .

Finally, these concepts are applicable to a de Broglie atom trap.  Instead of using a hollow fiber, Dowling\cite{dowling1} suggested using a bottle to trap the atoms.  The bottle has a narrow neck and a short bundle of narrow, hollow optical fibers are arranged side by side and fused to form a porous plug for sealing the neck of the bottle trap.  Blue detuned light can then be coupled from the walls of the bottle via contact coupling into the non-hollow parts of the plug in such a way that very little scattered light will exist in the hollow regions of the plug and the plug can effectively repel cold atoms which would otherwise come in contact with it.  Relatively warm atoms, having short de Broglie wavelengths, can then be injected directly through the small open areas of the fiber plug directly in to the bottle where the atoms are subsequently cooled using laser cooling with fiber couplers followed by evaporative cooling as has been described.  By cooling the atoms inside the bottle, the de Broglie wavelength will increase to the point where the atoms will be unable to pass through the small diameter hollow fibers in the porous plug and will now be trapped for considerable times inside the bottle to form an atom bottle trap.  The atom bottle geometry has the additional benefit that even lower temperatures can be achieved together with higher compression factors by going to a larger inner diameter of the bottle as compared to the hollow fiber.

\newpage

\begin{center}
{\large Figure Legends}
\end{center}

\noindent Fig.1. The combined effect of the repulsive barrier and vdW potential are shown here.  The barrier is repulsive in the sense that atoms moving toward the inner wall will be repelled back toward the origin for energies less than the barrier height.  For those atoms either tunneling through the barrier or otherwise getting very close to the inner wall, the strongly attractive van der Waals force dominates here as it tends to $-\infty$ and the atoms will be strongly drawn toward the inner wall and will quickly hit it.

\vspace*{0.1in}
\noindent
Fig.2. Negative potential energy well.  Note that the energy scale is about $10^4$ smaller than in Fig.1.  The very strong vdW potential tending to $-\infty$ is not shown here as the atom moves very close to the inner wall.  Specific parameters for this plot are: $a=10\mu$m, $\Delta =10^{10}$Hz, and $E_L =10^5$ V/m.

\vspace*{0.1in}
\noindent
Fig.3. Cross section through the repulsive barrier.  The energy scale here is about $10^4$ times too large to show the much smaller, negative energy potential well.

\vspace*{0.1in}
\noindent
Fig.4. Reduction in barrier to inner wall using 2 color total potential from eq. 3.  Here the barrier is reduced adiabatically by sweeping the frequency from $\Delta_{11}$ up to $\Delta_{14}$ or $\Delta_{15}$ where the  $\Delta_{1j}$ are in Hz and given in Fig.4.  The fixed values as used in eq. 3 are: $E_{1} =10^{\,5}$ V/m, $E_{2}= 10^{\,5}$ V/m, and $\Delta_{2} =-10^{\,8}$ Hz.  This has the potential to reduce the number of atoms lost in evaporative cooling.

\vspace*{0.1in}
\noindent
Fig.5.  Plot of $|\psi(\rho)|^2$, the probability density (unnormalized), and the negative potential energy well as a function of $\rho$.  The results of  numerically solving the Schr\"odinger eq. for the ground state of the negative potential energy well are shown here where $a =10.0\, \mu$m, $m_0=4.00 \times 10^{-26}$ kg, $SF=7.26 \times 10^{34}\,$J$^{-1}$, $E=-25668$,
$E_J=E/SF=3.54 \times 10^{-31}$ J, $T=2.57 \times 10^{-8}$ $^{o}$K, and $\rho_{match}=0.97$.  Note the differences in tunneling into the inner and outer potential walls:  there is significant tunneling to the left and less tunneling into the much steeper right hand side.  Specific parameters for this two-color odp potential are:
$a=10\mu$m, $\Delta_1 = 10^{10}$Hz, $\Delta_2 = -10^8$Hz,  $E_{L1} =10^5$ V/m, and $E_{L2}=10^{4.07}$V/m.

\newpage
\begin{center}{\bf Acknowledgments}\end{center}
One of us (H.S.P.) is pleased to acknowledge helpful discussions with the following:
\newline
D. Anderson, J. Dowling, S. Efros, G. Herling, M. Kasevich, M. Rosen, Y. Shih and L. Sica.

\newpage
\epsfxsize6.5in
\epsffile{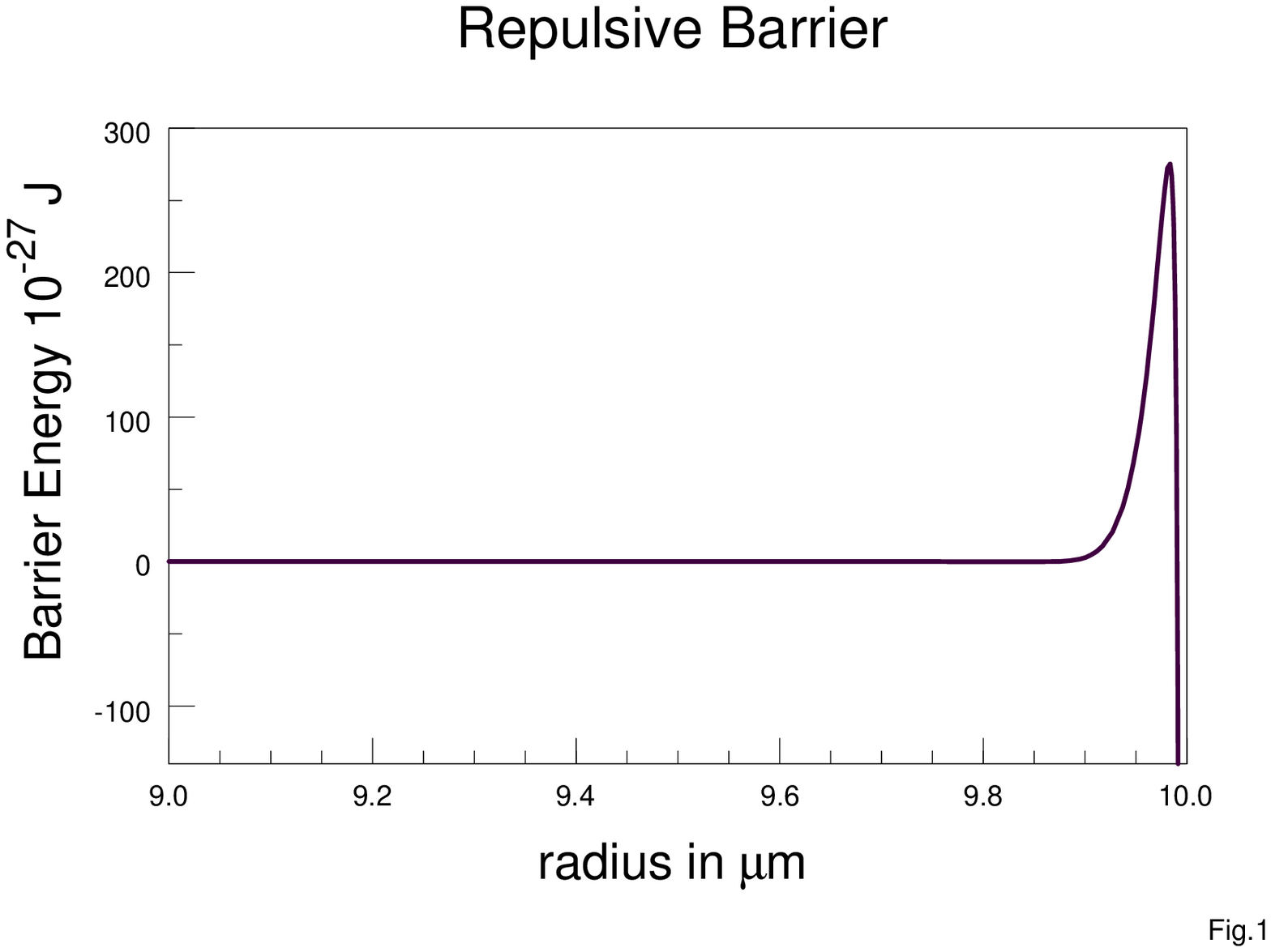}

\epsfxsize6.5in
\epsffile{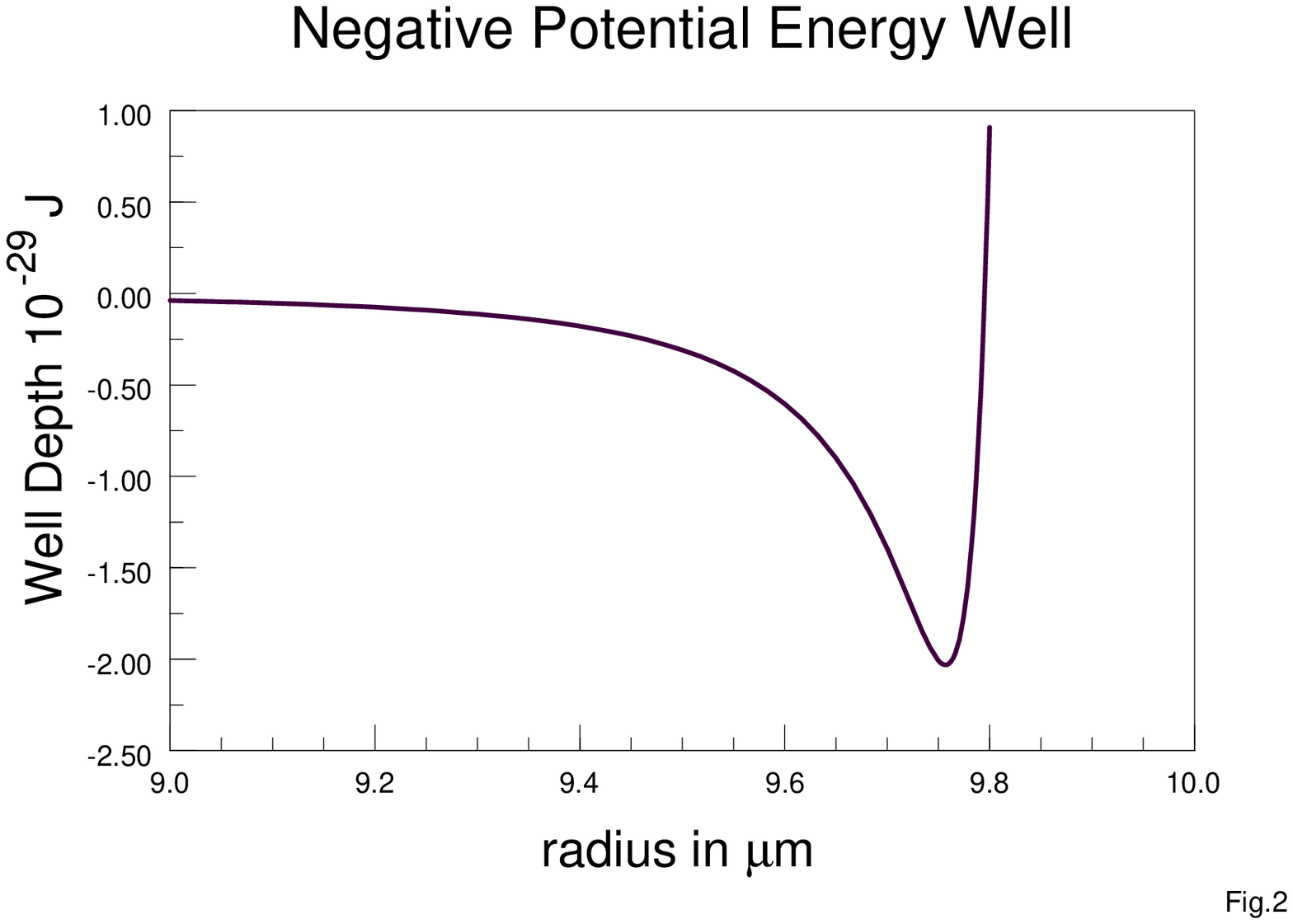}

\newpage
\epsfxsize6.5in
\epsffile{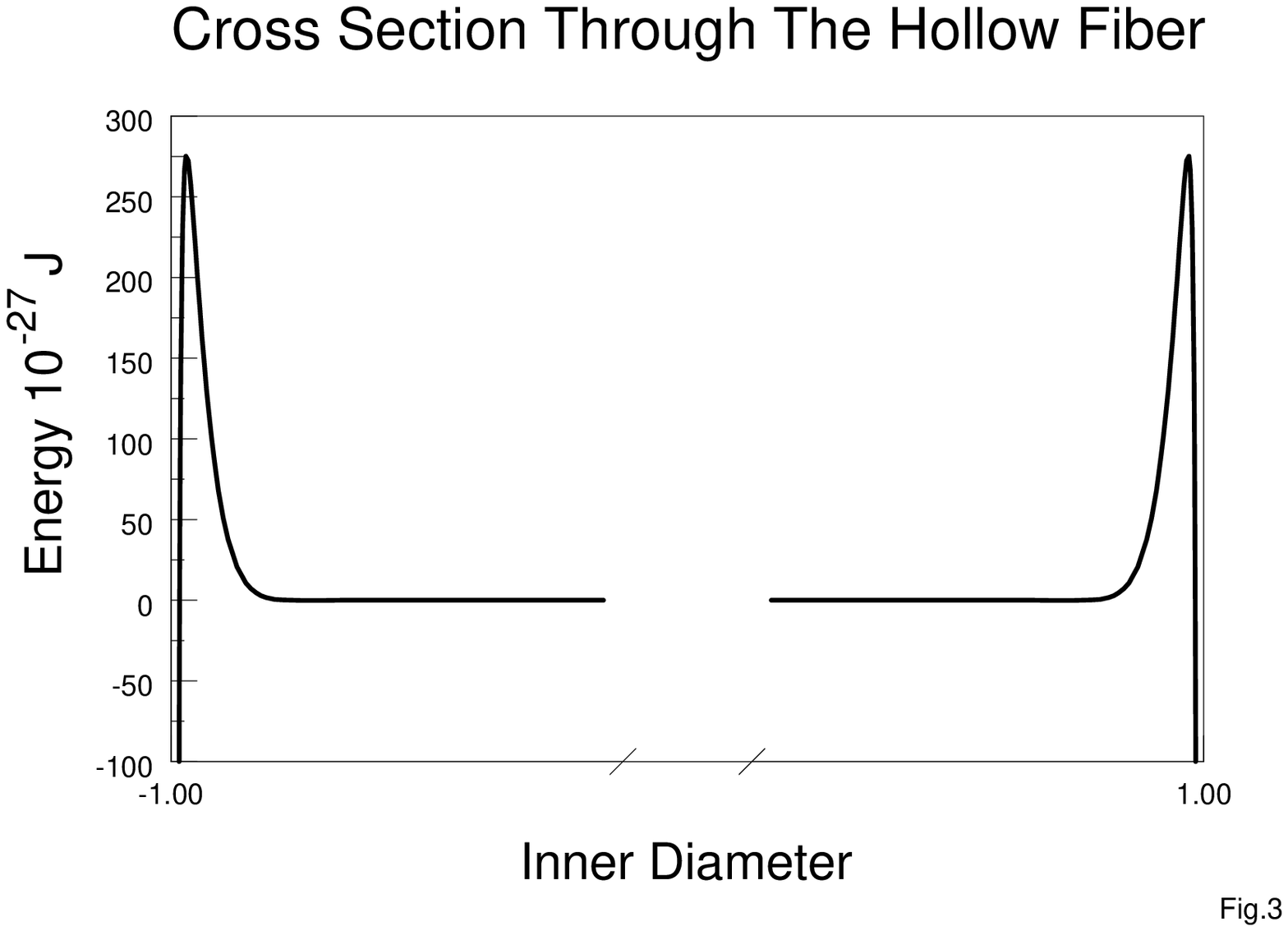}

\epsfxsize6.5in
\epsffile{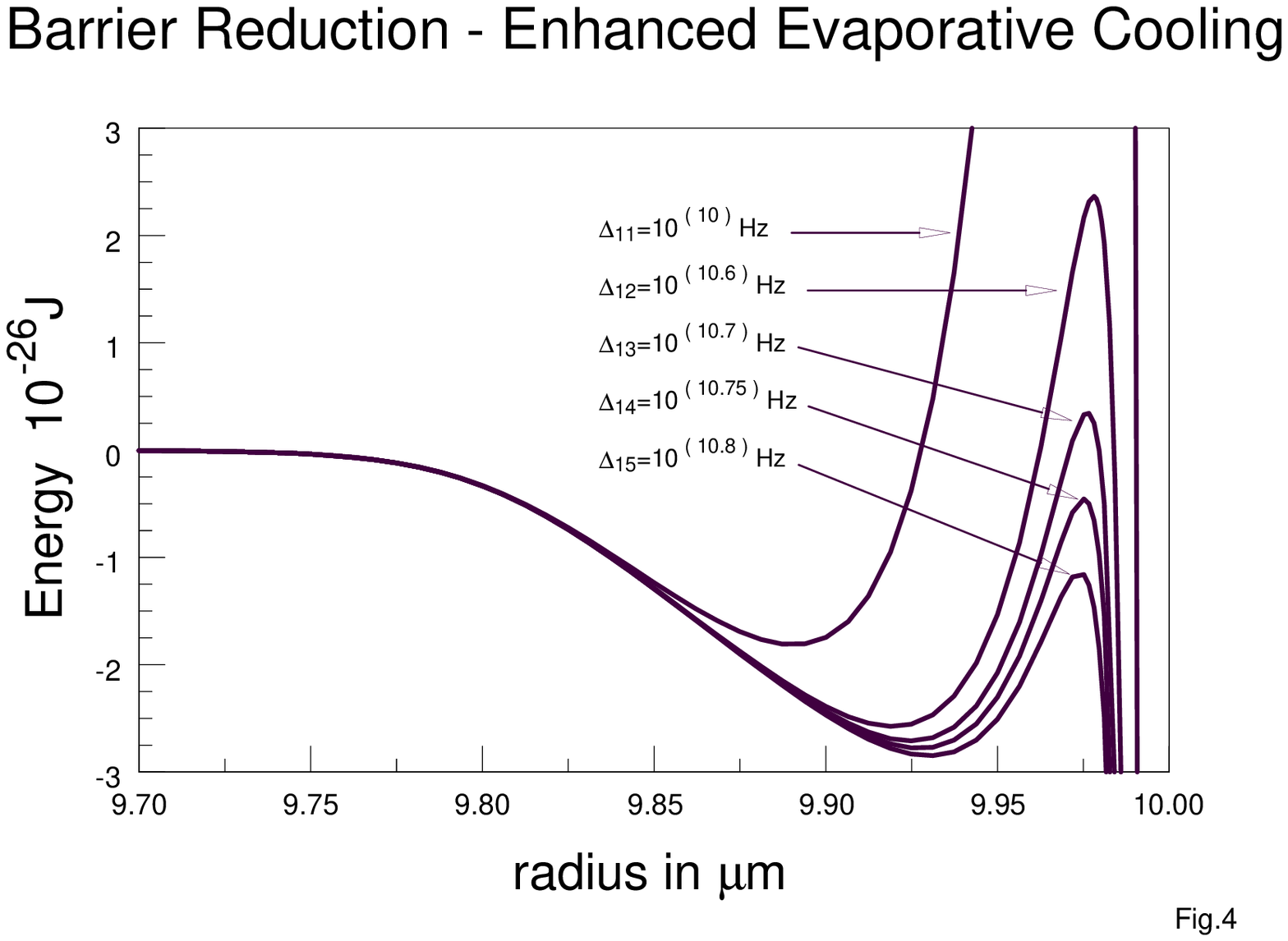}

\newpage
\epsfxsize6.5in
\epsffile{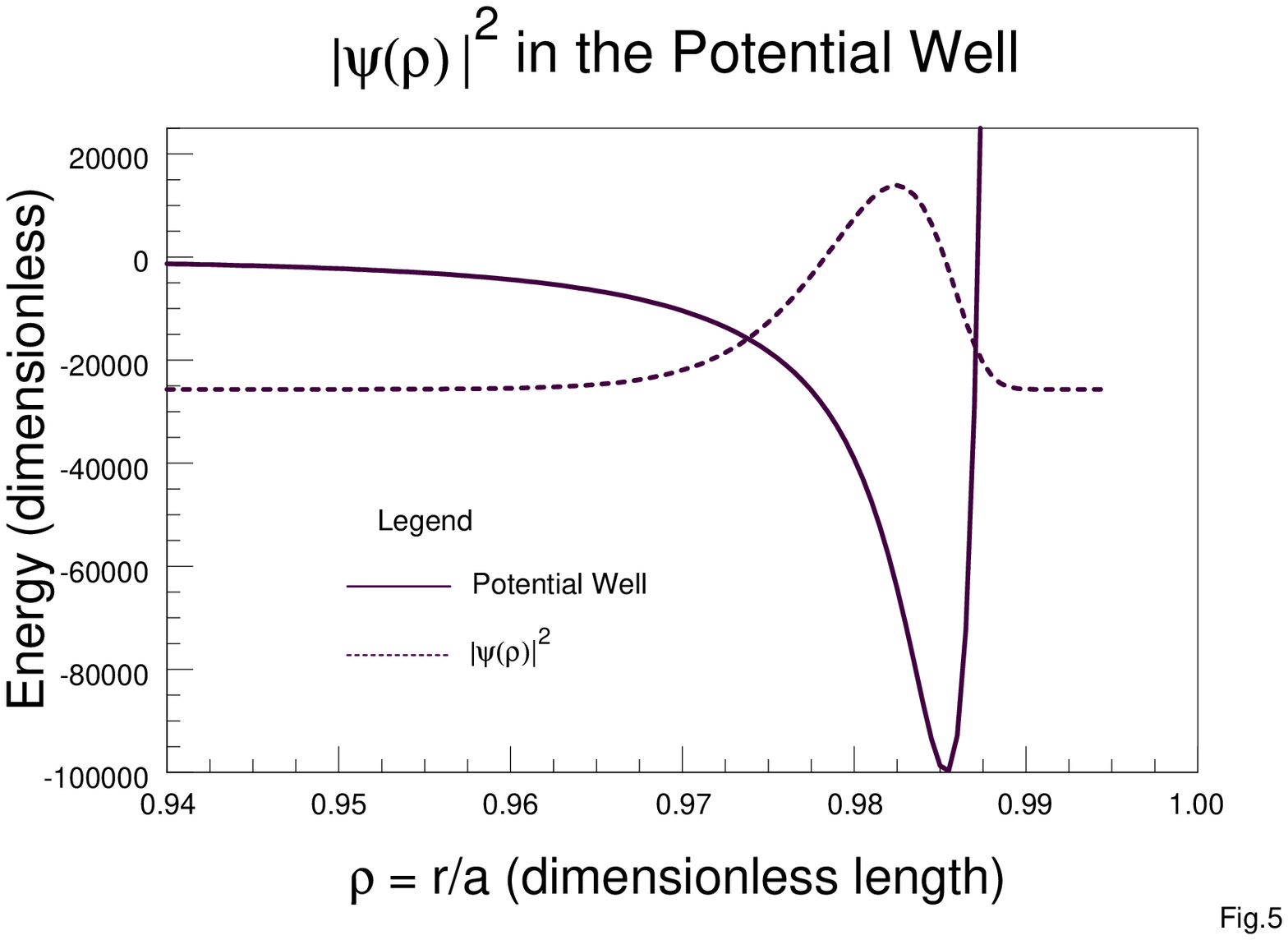}


\begin{thebibliography}{99}

\bibitem{cook} R.J. Cook and R.K. Hill, Opt. Commun. 43, 258 (1983).

\bibitem{marksteiner} S. Marksteiner, C.M. Savage, P. Zoller, and S.L. Rolston, Phys. Rev. A 50, 2680 (1994).

\bibitem{savage} C.M. Savage, S. Marksteiner, and P. Zoller, in Fundamentals of Quantum Optics III, edited by F. Ehlotzky,  Pub. Springer-Verlag, NY,NY (1993).

\bibitem{renn} M.J. Renn, E.A. Donley, E.A. Cornell, C.E. Wiemann, and D.Z. Anderson, Phys. Rev. A 53, 648 (1996).

\bibitem{ito} H. Ito, T. Nakata, K. Sakaki, M. Ohtsu, K.I. Lee, and W. Jhe., Phys. Rev. Lett. 76, 4500 (1996).

\bibitem{dowling} J. P. Dowling and J. Gea-Banacloche, Adv. At. Mol. Opt. Phys. 36, 1 (1996).

\bibitem{pilloff} H.S. Pilloff, Opt. Commun. 143, 25 (1997).

\bibitem{pilloff1} H.S. Pilloff, Opt. Commun. 179, 123 (2000).

\bibitem{masuhara} N. Masuhara, J.M. Doyle, J.C. Sandberg, D. Kleppner, T.J. Greytak, H.F. Hess, and G.P. Kochanski. Phys. Rev. Lett. 61,935 (1988).

\bibitem{ovchinnikov}Yu, B.Ovchinnikov, S.V. Shul'ga, and V.I. Balykin, Journal of Physics B, 24, 3173, (1991).

\bibitem{mabuchi}H. Mabuchi and H.J. Kimble, Optics Letters 19, 749 (1994).

\bibitem{Bohm}D. Bohm, Quantum Theory (Prentice Hall, New York, 1951, p. 70).

\bibitem{metcalf}H.J. Metcalf and P. van der Straton,Laser Cooling and Trapping (Springer-Verlag, New York, 1999, Chp. 12).

\bibitem{pilloff2}M.D. Pilloff and H.S. Pilloff, to be published.

\bibitem{dowling1}J.P. Dowling, Eds., M. G. Prentiss and W. D. Phillips, SPIE Proceedings 2995, Bellingham, WA, 126 (1997).

\end{thebibliography}
\end{document}